\documentclass[preprint,showpacs,preprintnumbers]{revtex4}
\usepackage{amssymb}
\usepackage{txfonts}
\usepackage{graphicx}
\usepackage{dcolumn}
\usepackage{bm}
\usepackage{subfigure}

\begin{document}

\title{Stability of vortex in a two-component superconductor}
\author{Jun-Ping Wang}
\affiliation{Department of Physics, Yantai University, Yantai 264005, P. R. China}

\begin{abstract}
Thermodynamic stability of composite vortex in a two-component
superconductor is investigated by the Ginzburg-Landau theory. The predicted
nature of these vortices has recently attracted much attention. Here we
consider axially symmetric quantized vortex and show that the stability of
vortex depends on three independent dimensionless parameters: $\kappa _{1}$,
$\kappa _{2}$, $\kappa _{\xi}$, where $\kappa _{i}(i=1,2)$ is the
Ginzburg-Landau parameter of individual component, $\kappa _{\xi }=\xi
_{1}/\xi _{2}$ is the ratio of two coherence lengths. We also show that
there exists thermodynamically stable vortex in type-1+type-2 or
type-2+type-2 materials over a range of these three parameters.
\end{abstract}

\maketitle

\bigskip

The existence of quantized vortex in a type-2 superconductor (SC) is one of
the most striking phenomena in condensed matter physics \cite{Abrikosov}.
The criterion for stability of vortex in a conventional SC is the
Ginzburg-Landau parameter $\kappa $, which is defined as the ratio of
penetration depth to coherence length: $\kappa =\lambda /\xi $ \cite%
{Ginzburg}. Vortex can exist as a thermodynamically favorable state under
external field in a type-2 material with $\kappa >1/\sqrt{2}$, while the
penetration of vortex is not thermodynamically favorable in a type-1 SC with
$\kappa <1/\sqrt{2}$.

\bigskip

Recently, there has been growing interest in investigating the vortex in
multi-component SCs \cite{koshelev,machida,babaev,goryo,1.5,VM2,Geurts}. The predicted nature of
vortex in these materials is quite different from that of vortex in a
conventional type-2 SC. Babaev and Speight showed that interaction potential
between two vortices in a two-component SC can be non-monotonic: intervortex
force is attractive at long range and repulsive at short range \cite%
{babaev}. The key question regarding Babaev and Speight's work is
whether vortices in a two-component SC are thermodynamically stable or not.
In this paper, we revisit this issue addressed in Ref. \cite{babaev}. The
idea is that, vortex can survive as a thermodynamically favorable state if
the Gibbs energy of the vortex state under the thermodynamic critical field
is smaller than that of the fully superconducting state (Meissner state). We
find that, the stability of vortex depends on three independent
dimensionless parameters: $\kappa _{1}$,$\kappa _{2}$,$\kappa _{\xi }$,
where $\kappa _{i}(i=1,2)$ is the Ginzburg-Landau parameter of individual
component, $\kappa _{\xi }=\xi _{1}/\xi _{2}$ is the ratio of two coherence
lengths. We also find that vortex is thermodynamically stable in a
type-1+type-2 or type-2+type-2 SC over a range of these three parameters.

\bigskip

Based on the Ginzburg-Landau model, we use free energy density in the
two-component SC as follows:

\begin{equation}
f=f_{n0}+\sum_{i=1}^{2}{\frac{\hbar ^{2}}{2m_{i}^{\ast }}|(\nabla -\frac{%
ie_{i}^{\ast }}{{\hbar }c}\mathbf{A})\Psi _{i}|^{2}}+V({|\Psi _{1,2}|}%
^{2})+\eta (\Psi _{1}^{\ast }\Psi _{2}+\Psi _{1}\Psi _{2}^{\ast })+\frac{1}{%
8\pi }(\nabla \times \mathbf{A})^{2},  \label{GL}
\end{equation}%
where $f_{n0}$ is the free energy density of the body in the normal state in
the absence of the magnetic field, $V({|\Psi _{i}|}^{2})=a_{i}{|\Psi _{i}|}%
^{2}+b_{i}{|\Psi _{i}|}^{4}/2\;(i=1,2)$. $\eta $ is a coefficient
characterizes Josephson coupling between two superconducting components. In
the following we consider in particular weak Josephson coupling limit and
set $\eta =0$. We also assume that the effective mass $m_{i}^{\ast }$ and
charge $e_{i}^{\ast }$ of two components are equal: $m_{i}^{\ast }=m^{\ast }$%
, $e_{i}^{\ast }=e^{\ast }$. There are four characteristic lengths: the
penetration depth $\lambda _{i}$ and coherence length $\xi _{i}$ for each
component are given by: $\lambda _{i}=(m^{\ast }c^{2}/4{\pi }e^{\ast 2}\Psi
_{i0}^{2})^{1/2}$, $\xi _{i}=\hbar /(2m^{\ast }|a_{i}|)^{1/2}$, where $\Psi
_{i0}=(-a_{i}/b_{i})^{1/2}$. The thermodynamic critical magnetic field of
the individual component is $H_{ct(i)}=\Phi _{0}/(2\sqrt{2}\pi \lambda
_{i}\xi _{i}),$ where $\Phi _{0}=hc/e^{\ast }$ is the flux quantum. The
magnetic field penetration depth and the thermodynamic critical magnetic
field of the system (\ref{GL}) are: $\lambda =(1/\lambda _{1}^{2}+1/\lambda
_{2}^{2})^{-1/2},\;H_{ct}=(H_{ct(1)}^{2}+H_{ct(2)}^{2})^{1/2}$. Note that $%
\lambda <\min (\lambda _{1},\lambda _{2})$, $H_{ct}>\max
(H_{ct(1)},H_{ct(2)})$.

\bigskip

\bigskip We consider axially symmetric quantized vortex in the model (\ref%
{GL}):

\begin{equation}
\Psi _{1}=\left\vert \Psi _{1}\right\vert e^{i\theta },\ \Psi
_{2}=\left\vert \Psi _{2}\right\vert e^{i\theta },\ \mathbf{A}=A(r)\mathbf{e}%
_{\theta }.  \label{vortex}
\end{equation}%
In order to study the stability of this vortex, we consider the Gibbs energy
difference between the vortex state under the thermodynamic critical field $%
H_{ct}$ and the fully superconducting state (Meissner state, or fully normal
state under the thermodynamic critical field since these must be equal)

\begin{equation}
\bigtriangleup G=G_{vortex}(H_{ct})-G_{M}=G_{vortex}(H_{ct})-G_{n}(H_{ct}).
\label{deltaG0}
\end{equation}%
Let us show that if $\bigtriangleup G<0$ then the isolated vortex can appear
as a thermodynamically favorable state under external field $H<H_{ct}$.
There are three possible states: fully superconducting state (Meissner
state), vortex state and fully normal state. And we note both the Gibbs
energies of vortex state and normal state are decreasing with increasing
field. If $\bigtriangleup G>0$, the normal state becomes energetically
favorable when the field exceeds the thermodynamic critical value and vortex
state is energetically unfavorable. If $\bigtriangleup G<0$, vortex state
becomes favorable under certain value of the external field which is smaller
than the thermodynamic critical field $H<H_{ct}$.

\bigskip

\bigskip The Gibbs energies of the vortex state and the Meissner state can
be written as
\begin{equation}
G_{vortex}(H_{ct})=\int d\mathbf{r}g_{v},\ G_{M}=G_{n}(H_{ct})=\int d\mathbf{%
r}g_{M}.  \label{2G}
\end{equation}%
Here $g_{v}=f_{v}-\mathbf{H}_{ct}\cdot (\mathbf{\nabla \times A})\mathbf{/}%
4\pi ,$\bigskip $g_{M}=f_{M}=f_{n0}-H_{ct}^{2}/8\pi $ are the Gibbs energy
densities of the vortex state and the Meissner state, respectively. To
investigate the vortex stability in the considered case, one can write down
\begin{equation}
\bigtriangleup G=\int d\mathbf{r}\left\{ \sum_{i=1}^{2}{\frac{\hbar ^{2}}{%
2m_{i}^{\ast }}|(\nabla -\frac{ie_{i}^{\ast }}{{\hbar }c}\mathbf{A})\Psi
_{i}|^{2}}+V({|\Psi _{1,2}|}^{2})+\frac{1}{8\pi }(\mathbf{H}_{ct}-\nabla
\times \mathbf{A})^{2}\right\} .  \label{deltaG00}
\end{equation}
We shall use instead of the variable $r$, the functions $\Psi _{i}$ and $%
\mathbf{A}$ the dimensionless quantities

\begin{equation}
\rho =\frac{r}{\lambda },\ \psi _{1}=\frac{\left\vert \Psi _{1}\right\vert }{%
\Psi _{10}},\ \psi _{2}=\frac{\left\vert \Psi _{2}\right\vert }{\Psi _{20}}%
,\ A=\frac{\left\vert \mathbf{A}\right\vert }{H_{ct}\lambda }.
\label{dimensionless}
\end{equation}%
In the following we calculate $\Delta G$ and find

\begin{equation}
\Delta G=\frac{H_{ct}^{2}\lambda^{2}}{4}\int\nolimits_{0}^{\infty }\rho
d\rho \left\{ \sum_{i=1}^{2}\frac{C_{i}}{B_{i}}[2A_{i}(\frac{d\psi _{i}}{%
d\rho })^{2}+(\sqrt{B_{i}}\psi _{i}A-\sqrt{2A_{i}}\frac{\psi _{i}}{\rho }%
)^{2}-2\psi _{i}^{2}+\psi _{i}^{4}]+\left( 1-\frac{1}{\rho }\frac{\partial }{%
\partial \rho }(\rho A)\right) ^{2}\right\} .  \label{deltaG}
\end{equation}%
Here $A_{1}=1/\kappa _{1}^{2}+\kappa _{\xi }^{2}/\kappa
_{2}^{2},\;B_{1}=(\kappa _{2}^{2}+\kappa _{1}^{2}\kappa _{\xi }^{4})/(\kappa
_{2}^{2}+\kappa _{1}^{2}\kappa _{\xi }^{2}),\;C_{1}=\kappa _{2}^{2}/(\kappa
_{2}^{2}+\kappa _{1}^{2}\kappa _{\xi }^{2}),\;A_{2}=1/\kappa
_{2}^{2}+1/\kappa _{1}^{2}\kappa _{\xi }^{2},\;\;B_{2}=(\kappa
_{2}^{2}+\kappa _{1}^{2}\kappa _{\xi }^{4})/[\kappa _{\xi }^{2}(\kappa
_{2}^{2}+\kappa _{1}^{2}\kappa _{\xi }^{2})],\;C_{2}=\kappa _{1}^{2}\kappa
_{\xi }^{2}/(\kappa _{2}^{2}+\kappa _{1}^{2}\kappa _{\xi }^{2})$, $\kappa
_{i}=\lambda _{i}/\xi _{i}(i=1,2)$ is the Ginzburg-Landau parameter of
individual component, $\kappa _{\xi }=\xi _{1}/\xi _{2}$ is the ratio of two
coherence lengths. The Ginzburg-Landau equations which determine the profile
of the vortex solution are determined by minimizing the $\Delta G$ with
respect to functions $\psi _{i}(i=1,2)$ and $A$%
\[
A_{1}(\psi _{1}^{^{\prime \prime }}+\frac{\psi _{1}^{^{\prime }}}{\rho }-%
\frac{\psi _{1}}{\rho ^{2}})+\sqrt{2A_{1}B_{1}}\frac{\psi _{1}}{\rho }A-%
\frac{1}{2}B_{1}A^{2}\psi _{1}+\psi _{1}-\psi _{1}^{3}=0,
\]%
\[
A_{2}(\psi _{2}^{^{\prime \prime }}+\frac{\psi _{2}^{^{\prime }}}{\rho }-%
\frac{\psi _{2}}{\rho ^{2}})+\sqrt{2A_{2}B_{2}}\frac{\psi _{2}}{\rho }A-%
\frac{1}{2}B_{2}A^{2}\psi _{2}+\psi _{2}-\psi _{2}^{3}=0,
\]%
\begin{equation}
A^{^{\prime \prime }}+\frac{A^{^{\prime }}}{\rho }-\frac{A}{\rho ^{2}}%
=(C_{1}\psi _{1}^{2}+C_{2}\psi _{2}^{2})A-(C_{1}\sqrt{\frac{2A_{1}}{B_{1}}}%
\frac{\psi _{1}^{2}}{\rho }+C_{2}\sqrt{\frac{2A_{2}}{B_{2}}}\frac{\psi
_{2}^{2}}{\rho }).  \label{GLeqs}
\end{equation}%
The order parameters vanish in the vortex core which is the phase
singularity of both components. And it can be expected that the potential ${A%
}$ correlates linearly with $\rho$ in the core region due to the constant
value of the field inside the core. Then the boundary conditions at $\rho=0$ are

\begin{equation}
\psi _{1}(0)=\psi _{2}(0)=A(0)=0.  \label{b1}
\end{equation}%
We assume the following power series solutions to Eqs. (\ref{GLeqs}) in the
region of the core: $\psi _{1}(\rho )=\sum_{n=0}^{\infty }b_{n}\rho ^{n},\
\psi _{2}(\rho )=\sum_{n=0}^{\infty }c_{n}\rho ^{n},\ A=\sum_{n=0}^{\infty
}a_{n}\rho ^{n}$ and can prove that

\[
\psi _{1}=b_{1}\rho -\frac{1}{8A_{1}}(\sqrt{2A_{1}B_{1}}a_{1}+1)b_{1}\rho
^{3}+O(\rho ^{5}),\ \psi _{2}=c_{1}\rho -\frac{1}{8A_{2}}(\sqrt{2A_{2}{B_{2}}%
}a_{1}+1)c_{1}\rho ^{3}+O(\rho ^{5}),\
\]

\begin{equation}
A=a_{1}\rho -\frac{1}{8}(C_{1}\sqrt{\frac{2A_{1}}{B_{1}}}b_{1}^{2}+C_{2}%
\sqrt{\frac{2A_{2}}{B_{2}}}c_{1}^{2})\rho ^{3}+O(\rho ^{5}).  \label{Taylor1}
\end{equation}%
Here $a_{1},b_{1},c_{1}$ are three constants which will be deduced from the
solutions far from the vortex core \cite{far}:

\begin{equation}
\psi _{1}(\infty )=\psi _{2}(\infty )=1,\ B\left( \infty \right) =(\frac{1}{%
\rho }\frac{\partial }{\partial \rho }(\rho A))|_{\infty }=0,  \label{b2}
\end{equation}%
where $B=({1/\rho )\partial (A\rho )/\partial \rho }$ is the scaled magnetic
field: $B=\left\vert \nabla \times \mathbf{A}\right\vert /H_{ct}$. It is
clear from (\ref{deltaG}), (\ref{GLeqs}), (\ref{b1}) and (\ref{b2}) that the
sign of $\Delta G$, and the stability of vortex in the model are determined
by the three independent dimensionless parameters: $\kappa _{1}$,$\kappa _{2}
$,$\kappa _{\xi }$.

\begin{figure}[tbp]
\begin{center}
\includegraphics[width=\textwidth,angle=0]{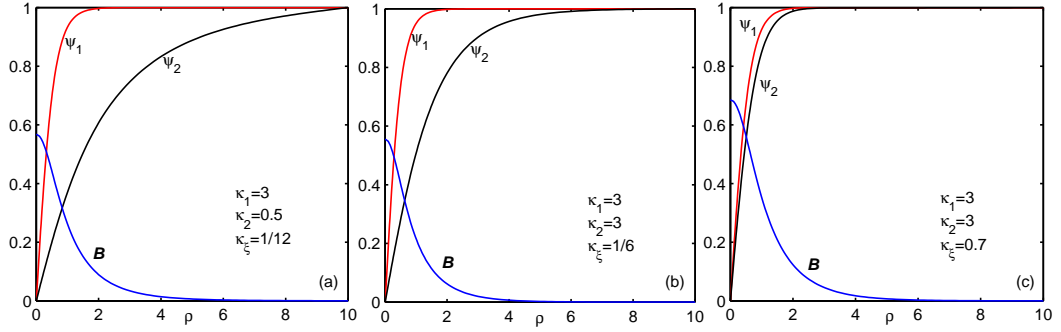}
\end{center}
\caption{Stable vortex solutions in a two-component SC. (a) is a vortex
solution in a type-1+type-2 SC, while (b) and (c) are vortex solutions in
the type-2+type-2 SCs. Note the similarity of the vortex configurations in
(a) and (b).}
\label{fig:figu1}
\end{figure}

\begin{table}[tbp]
\caption{Stable vortex solutions in a two-component SC}
\label{tab:mytab1}\centering
\par
\tabcolsep=12pt\centering
\smallskip
\par
\begin{tabular*}{128mm}{ccccccc}
\hline\hline
$\kappa_{1}$ & $\kappa_{2}$ & $\kappa_{\xi}$ & $b_{1}$ & $c_{1}$ & $a_{1}$ &
$\bigtriangleup G/({H_{ct}^{2}\lambda^{2}}/{4})$ \\ \hline
3 & 0.5 & 1/12 & 1.7253 & 0.4111 & 0.2830 & -0.9044 \\
3 & 3 & 1/6 & 1.8973 & 0.5843 & 0.2762 & -0.5721 \\
3 & 3 & 0.7 & 1.6275 & 1.2383 & 0.3425 & -1.1159 \\ \hline\hline
\end{tabular*}%
\end{table}

We have numerically solved Ginzburg-Landau eqs. (\ref{GLeqs}) with boundary
conditions (\ref{Taylor1}),(\ref{b2}) and identified thermodynamically
stable vortex solutions in the type-1+type-2 and type-2+type-2 SCs, as
predicted previously by the surface energy calculations \cite{surface}.
Figure 1 illustrates several examples of stable vortex solutions. We found
that the vortex stability in a two-component SC depends not only on the
Ginzburg-Landau parameter of individual component, but also on the third
parameter $\kappa _{\xi }={\xi_{1}}/{\xi_{2}}$.

\bigskip

Here we want to relate works in Ref. \cite{babaev} to the results of the
present work. In their paper, Babaev and Speight identified three
characteristic lengths in the model (\ref{GL}): penetration depth $\lambda $
and coherence lengths of two components $\xi _{1}$,$\xi _{2}$. They
presented an example of vortex solution with $\xi _{1}/\lambda =\sqrt{2}/3$,
$\xi _{2}/\lambda =4\sqrt{2}$. Our results reveal that the intrinsic
parameters which determine the magnetic properties of a two-component SC are
three independent dimensionless parameters: $\kappa _{1}$,$\kappa _{2}$,$%
\kappa _{\xi }$. It is easy to verify that the values of these three
parameters are not unique for given $\xi _{1}/\lambda $ and $\xi
_{2}/\lambda $. This means that, the vortex solution presented in Ref. \cite%
{babaev} is \textit{not} the unique solution. In figure 2 we present several
examples in which $\kappa _{1}$,$\kappa _{2}$ are different while $\xi
_{1}/\lambda =\sqrt{2}/3$, $\xi _{2}/\lambda =4\sqrt{2}$. It is clear from
table 2 that the slopes of $\psi _{1},\psi _{2}$ and $A$ near the center of
vortex core are slightly different. This proved that these vortex solutions
are different solutions of Ginzburg-Landau eqs. (\ref{GLeqs}). In
particular, there are cases in which two component are both of type-2 ((b)
and (c) in the figure 2), which were not mentioned in Ref. \cite{babaev}.

\bigskip

\begin{figure}[tbp]
\begin{center}
\includegraphics[width=1.0\textwidth,angle=0]{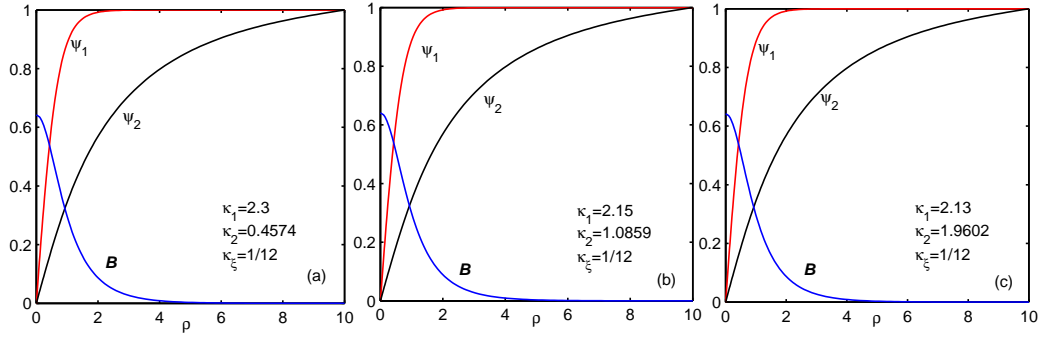}
\end{center}
\caption{Examples of stable vortex solutions with $\protect\xi_{1}/\protect%
\lambda=\protect\sqrt2/3$ and $\protect\xi_{2}/\protect\lambda=4\protect\sqrt%
2$. (a) is a vortex solution in a type-1+type-2 SC, while (b) and (c) are
vortex solutions in the type-2+type-2 SCs. Detailed parameters can be found
in the table 2 below.}
\label{fig:figu2}
\end{figure}

\begin{table}[tbp]
\caption{Stable vortex solutions with $\protect\xi_{1}/\protect\lambda=%
\protect\sqrt{2}/3$ and $\protect\xi_{2}/\protect\lambda=4\protect\sqrt{2}$}
\label{tab:mytab2}\centering
\par
\tabcolsep=12pt\centering
\smallskip
\par
\begin{tabular*}{167mm}{ccccccccc}
\hline\hline
$\kappa_{1}$ & $\kappa_{2}$ & $\kappa_{\xi}$ & $\xi_{1}/\lambda$ & $%
\xi_{2}/\lambda$ & $b_{1}$ & $c_{1}$ & $a_{1}$ & $\bigtriangleup G/({%
H_{ct}^{2}\lambda^{2}}/{4})$ \\ \hline
2.3 & 0.4574 & 1/12 & $\sqrt{2}/3$ & $4\sqrt{2}$ & 1.4096 & 0.3753 & 0.3161
& -0.9122 \\
2.15 & 1.0859 & 1/12 & $\sqrt{2}/3$ & $4\sqrt{2}$ & 1.4194 & 0.3813 & 0.3197
& -0.6696 \\
2.13 & 1.9602 & 1/12 & $\sqrt{2}/3$ & $4\sqrt{2}$ & 1.4208 & 0.3821 & 0.3203
& -0.6387 \\ \hline\hline
\end{tabular*}%
\end{table}

A notable aspect of our results is that in a type-1+type-2 SC, stable vortex
solution always has a extended core associated with the type-1 component
(see fig.1(a),fig.2(a)). It is pointed that non-monotonic interaction
between vortices originates from this exceptional vortex configuration \cite%
{babaev,1.5}. However, in a type-2+type-2 SC, the situation is much
more complicated. We found that, stable vortex solutions in a type-2+type-2
SC may have a extended core (whose range is much larger than that of
penetration depth of the system, see fig. 1 (b),fig.2(b),2(c)), or may have
contracting core (whose range is smaller than the penetration depth, see
fig. 1(c)). This means that, in a type-2+type-2 material, intervortex fore
may be attractive at long range and repulsive at short range, as that of
vortices in a type-1+type-2 SC. Alternatively, intervortex force in a
type-2+type-2 SC may be repulsive at all range, as that of vortices in a
conventional type-2 SC.

\bigskip

In conclusion, we have identified the intrinsic parameters which determine
the stability of vortex in a two-component SC. Isolated vortex
can appear as a thermodynamically stable state in a type-1+type-2 or
type-2+type-2 SC over a range of these three parameters: $\kappa_{1},\kappa_{2},
\kappa_{\xi}$.

\bigskip

This work was supported by the National Natural Science Foundation of China
(No. 10547137).


\begin{references}
\bibitem{far} Here an alternative boundary condition for ${A}$ at infinity is $A(\infty)=\frac{\sqrt{2}}{\rho(\infty)} \sqrt{\frac{A_{1}}{B_{1}}}=\frac{\sqrt{2}}{\rho(\infty)} \sqrt{\frac{A_{2}}{B_{2}}}$, based on the fact that the vortices considered in the present work carry unit flux quanta $\Phi _{0}=hc/e^{\ast }$. However, eqs. (\ref{b2}) show a better convergence in the numerical methods we used.
\end{references}
\end{document}